# Artificial neural network based chemical mechanisms for computationally efficient modeling of kerosene combustion

Jian An[a,b], Guo Qiang He[a], Kai Hong Luo[b], Fei Qin[a*], Bing Liu[a]

[a] Science and Technology on Combustion, Internal Flow and Thermal-structure Laboratory, Northwestern Polytechnical University, Xi'an Shaanxi 710072, PR China

[b] Department of Mechanical Engineering, University College London, London WC1E 7JE, United Kingdom

**Corresponding Authors[*]:** Fei Qin     E-mail: qinfei@nwpu.edu.cn

## Abstract

To effectively simulate the combustion of hydrocarbon-fueled supersonic engines, such as rocket-based combined cycle (RBCC) engines, a detailed mechanism for chemistry is usually required but computationally prohibitive. In order to accelerate chemistry calculation, an artificial neural network (ANN) based methodology was introduced in this study. This methodology consists of two different layers: self-organizing map (SOM) and back-propagation neural network (BPNN). The SOM is for clustering the dataset into subsets to reduce the nonlinearity, while the BPNN is for regression for each subset. The entire methodology was subsequently employed to establish a skeleton mechanism of kerosene combustion with 41 species. The training data was generated by RANS simulations of the RBCC combustion chamber, and then fed into the SOM-BPNN with six different topologies (three different SOM topologies and two different BPNN topologies). By comparing the predicted results of six cases with those of the conventional ODE solver, it is found that if the topology is properly designed, high-precision results in terms of ignition, quenching and mass fraction prediction can be achieved. As for efficiency, 8~ 20 times speedup of the chemical system integration was achieved, indicating that it has great potential for application in complex chemical mechanisms for a variety of fuels.

Rocket-based combined cycle (RBCC)

# 1. Introduction

Rocket-based combined cycle (RBCC) engines seamlessly combine the advantages of high thrust/weight ratio of rocket engines and high specific impulse of jet engines to effectively operate at multiple modes ranging from take-off, high altitude hypersonic cruise, to orbit [1]. Its highly integrated design and superior performance make it most likely to be used in the next generation of hypersonic RLV, and one of the most promising propulsion systems for Single Stage To Orbit (SSTO). With the deepening of research on combustion process and the increasing demand for efficient design, high-fidelity numerical simulation has attracted more and more attention due to the difficulty and high cost of experiments. However, to effectively simulate the combustion of these hydrocarbon-fueled engines, a detailed mechanism for chemistry is usually required. The mechanism mathematically represents a large set of nonlinear stiff ordinary differential equations (ODEs) and must be integrated over every spatial point and every time step, hence, the computational cost of such ODEs in unsteady, three-dimensional reacting flows is prohibitive, even as a high-performance computing system have become accessible within the last decade. There are several ways to accelerate the calculation. Using reduced mechanisms is a widely used method to overcome this limitation. Examples are the methodologies for generating pre-reduced skeleton mechanisms, such as the multi-generation path flux analysis (PFA) method [2], the direct relation graph (DRG) method [3], or the DRG with error propagation (DRGEP) method [4], and approaches for generating local mechanisms for each grid and time step based on local thermodynamic conditions, namely the dynamic adaptive chemistry (DAC) method [5, 6].

Under the assumption that the flow field and chemistry are loosely coupled, the thermochemical states can be pre-computed or stored for reuse in real time, thus reducing the expensive real time calculations in the simulation process.

A straightforward approach is the lookup table (LUT) method proposed by Chen et al. [7], which directly stores thermodynamic states and reuses them. A pivotal approach for tabulation is the in situ adaptive tabulation (ISAT) [8], where the table of thermochemical states is generated on-the-fly and reused in later time steps using a binary tree algorithm. Although many efforts, such as the most recently used (MRU) and dynamic pruning strategy (DP) [9], have been made to improve the efficiency of the modification and retrieve of the table, the performance of ISAT would decrease significantly when dealing with large-scale unsteady numerical simulations like LES due to the extremely big table required [6].

Following this avenue, a good choice to overcome the shortcomings of chemistry tabulation is to use artificial neural networks (ANNs) to store thermochemical states. ANNs can efficiently represent the non-linear relationship between the input and output data, such as changes in the thermodynamic space. Chen et al. [10] took the early attempts to apply ANNs to represent chemical reactions, using a three-step chemistry mechanism for hydrogen turbulent combustion. The runtime and memory storage requirements were significantly reduced compared to the traditional look-up table technique as well as the direct integration approaches.

Subsequently, a reduced mechanism contained 4 steps and 8 species ($CH_4$, $O_2$, $CO_2$, $H_2O$, $CO$, $H_2$, $H$ and $N_2$) was employed as the target of ANN. By comparing the results of direct integration with those of a large number of random samples, a good agreement is obtained [11]. Blasco et al. [12] applied the ANNs for a reduced, 5 steps and 8 species mechanism for methane–air combustion, which is derived from the full GRI 2.11 mechanism. In this work, they proposed to divide the samples of thermodynamics space into sub-domains and use multiple ANNs with the aim to get a dedicated ANN for each sub domain, resulting a two-layer topology network. In first layer, self-organizing map (SOM), a technique that can divide high dimensional data into several pattern by calculating the Euclidean distance, was applied for the automatic partitioning. Combined with the ANNs for representation, the SOM-ANN topology was established and tested by simulating a Partial Stirred Reactor (PaSR). The presented approach was validated and encouraging results were reported.

Furthermore, Chatzopoulos and Rigopoulos [13] and Franke et al. [14] developed a method of combining the Rate-Controlled Constrained Equilibrium (RCCE) with the SOM-ANN for a mechanism of CH4–air combustion with 16 species. The training datasets were generated from an abstract problem based on the laminar flamelet equation and consist of the thermodynamics states with different mixture fractions and strain rates. After training, the simulation results show that RCCE-SOM-ANN topology can reduce the CPU time by about two orders of magnitude with excellent accuracy.

In the above work, the target mechanism for ANN is relatively small (up to 16 species to our knowledge) and is for simple fuels like hydrogen or methane. However, most of the fuels used in engineering applications are complex hydrocarbon fuels, and their mechanisms usually involve hundreds to thousands of species. In addition, most of the trained ANN models are verified using model flames, such as Sandia Flames or Sydney Flames, and their performance in the actual situation still needs to be tested, especially in the case of supersonic combustion. Therefore, in this study, we are committed to constructing an ANN-based mechanism of kerosene for supersonic turbulent combustion and verifying it in a practical RBCC combustion chamber. The remainder of this paper is organized as follows. Section 2 introduces the new methodology involved in this study. Section 3 briefly describes test cases and the numerical approach. Section 4 presents performance analyses conducted of DAC-ST and DAC-DT. Section 5 concludes this paper.

## 2. Methodology

### 2.1 Overview

In the simulation of a reactive flow with the Strang-based splitting scheme, the chemical alteration is separated from that of other physical processes, therefore the change of thermodynamics states is governed by a set of nonlinear ODEs, which can be expressed as:

$$\frac{d\varphi}{dt} = S(\varphi) \tag{1}$$

where $\varphi$ is the vector of the thermo-chemical composition, i.e. pressure $p$, temperature $T$, and the vector $Y$ of species mass fractions; and $S$ is its rate of change due to chemical reactions. To determine the thermochemical composition altered by chemical reactions over a time step, the CFD solver loops over all the computational cells and then calculates the integration of each reaction step, typically using stiff ODE solvers. To determine the thermochemical composition of a chemical reaction that changes over a time step, the integrator need to loop over all the computational cells. However, the thermochemical states in many cells are similar, and these repeated calculations consume a lot of computational resources, especially for a mechanism of kerosene. Therefore, to speed up the integration, an effective way is to store and reuse the map of the change of thermochemical states.

A two-layer structure as shown in Fig. 1, SOM-BPNN, was deployed in this work, in which SOM was responsible for classifying datasets because otherwise it would be difficult to obtain satisfactory model by using only one ANN. Furthermore, the back-propagation artificial neural network (BPNN), one type of ANN, was used to regress each subclass.

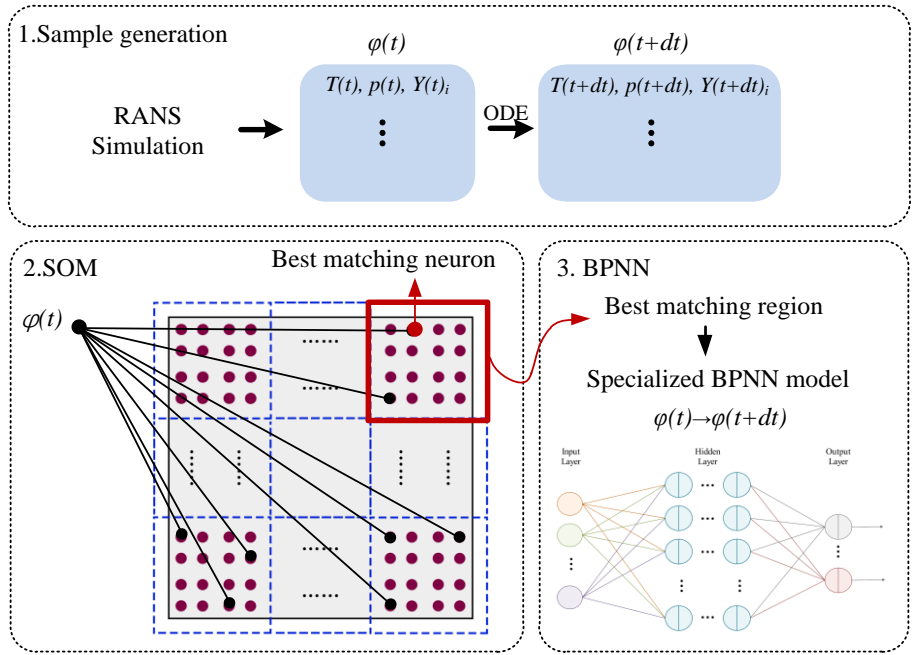

**Fig. 1.** Schematic of the training process of the SOM- BPNN topology.

**2.2 Sample generation**

The sample generation is the first step. By learning the samples, the neural network can approximate the complex nonlinear system, that is, the applicability of the obtained model depends on the training samples. For any chemical reaction, the thermodynamic state at time *t*, $\varphi(t)$, and that after a time step, $\varphi(t+\Delta t)$, are a pair of samples. When a large number of samples through a pre-calculation were collected to form a dataset, it can represent the entire process of chemical integration. In this study, the dataset was generated by a pre-simulated case as detailed in section 3.1.

**2.3 The SOM concept and implementation**

Represented by the dataset, the change of the thermochemical state driven by the chemical reaction is a high-dimensional system with complex nonlinear behavior. Although it is theoretically possible to employ only a single ANN to deal with such a problem, it is not a good choice because of poor computational efficiency and accuracy. Therefore, the SOM technique was introduced to divide the data set into several clusters [15]. The SOM is based on unsupervised learning, typically employed to perform pattern classification tasks. For the training process, once the samples are fed into the SOM, the algorithm adjusts the position of each neuron to minimize the Euclidean distance from more input samples. After the training phase, each neuron is the center of some input, representing a part of the input space. Meanwhile, the sum of distance of all samples to their center can be quantified by Eq. (2), the quantization error, where $x_i$ and $m_{i-best}$ represent the $i$-th sample and its best match neuron. The smaller the Q error, the more refined the classification. Finally, by finding the neuron with the smallest Euclidean distance from the input sample, the trained self-organizing tree model can assign the new sample to the subdomain.

$$Q = \sum_i \left\| x_i - m_{i-best} \right\|^2 \quad (2)$$

**2.4 The BPNN concept and training**

Through the previous step, the sample set was divided into subsets, and each subset has a BPNN for regression. BPNN is a kind of ANN widely used in regression analysis, which can deal with non-linear relations efficiently [16]. It usually has three parts: one input layer, one or more hidden layers and one output layer; and each of these layers has several neurons, which are mathematically represented by a weight matrix and a bias matrix. However, the weights and biases are randomly initialized at the beginning of training, which makes the training tricky. Therefore, the particle swarm optimization (PSO) method is employed in choosing a better initial value in this study because the BPNN-PSO has excellent performance in accelerating convergence [16]. The PSO is a population-based optimization method that simulates the social behavior of birds. This method adopts the method of population to search and finds the optimal solution through the cooperation and information sharing among individuals in the group, which enables it to search more regions in the solution space of the objective function to be optimized at the same time. A complete algorithm can be found in Refs [17, 18]. After initialization, the Adam optimizer [19], which is an efficient algorithm, is used to train each BPNN. When this is done, a complete SOM-BPNN-based mechanism is established and ready to use.

## 3. Application of SOM-BPNN method in a kerosene-fueled supersonic combustion

In this section, the SOM-BPNN was applied to represent a kerosene (represented by $C_{10}H_{22}$) mechanism with 41 species and 132 elemental reactions [20]. This skeletal mechanism is derived from a detailed kerosene high temperature mechanism of $C_{10}H_{22}$ consisting of 121 species and 866 reactions [21] and has been verified to have good accuracy [20]. As shown in Fig. 2, The comparisons of flame temperature, ignition delay time, ignition process and extinction predicted by the skeletal mechanism show an overall good accordance with the detailed mechanism [20].

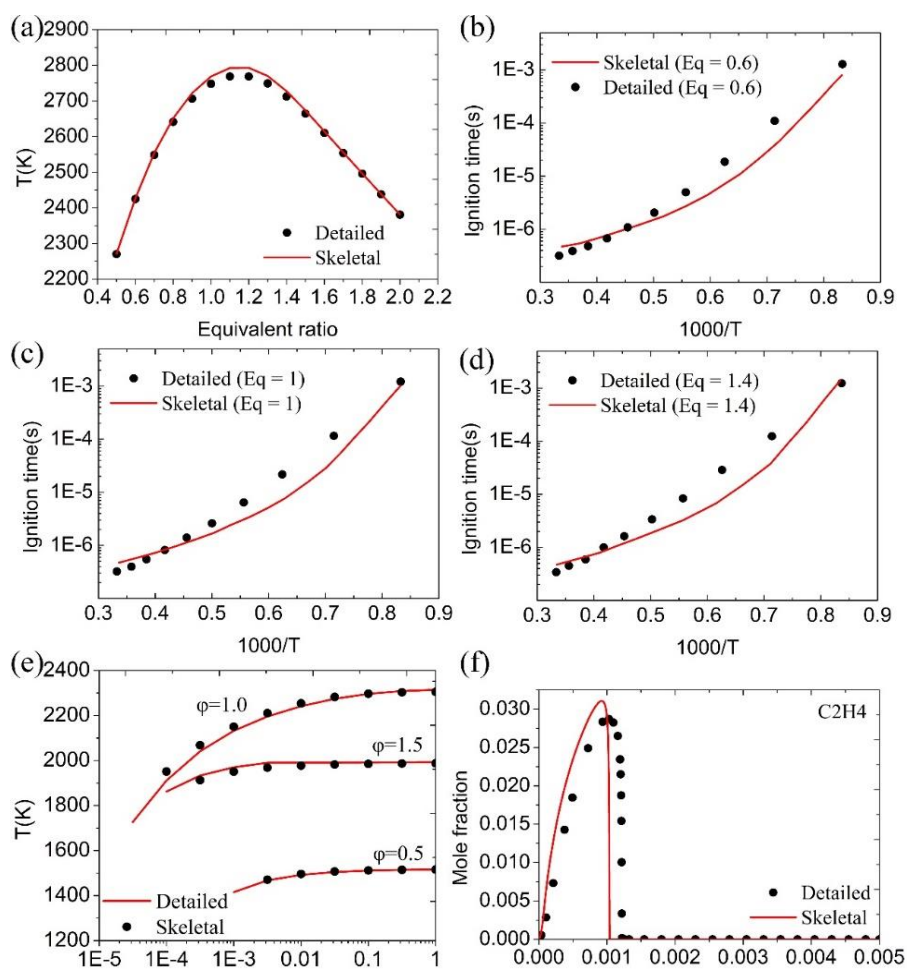

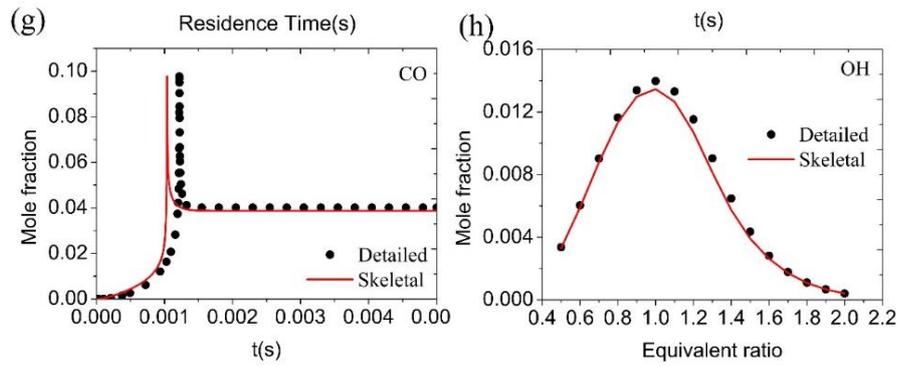

**Fig. 2.** Comparisons between the skeletal mechanism and the detailed mechanism.

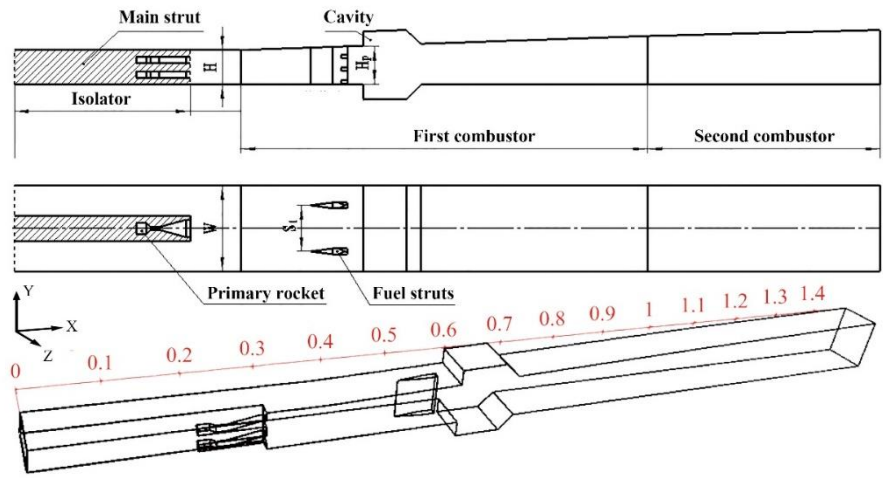

**Fig. 3.** Configuration of the RBCC engine combustor.

**Table 2** Simulation conditions.

| Inlet | | Rocket Inlet | | Second-fuel |
|---|---|---|---|---|
| Temperature, K | Mass, g/s | Temperature, K | Mass, g/s | Equivalence ratio |
| 1300 | 4000 | 1900 | 120 | 0.7 |

As described in Section 2.2, the pre-calculated case for sample generation is a RANS simulation of a full-scale RBCC engine corresponding to the flight condition at Ma=5.5 [20]. As shown in Fig. 3, it has a main strut, an isolator, a primary rocket, two struts for secondary fuel injection and flame stabilization, two cavity flame holders and multi-stage rectangular combustors. The main strut provides the Mach number required for the inlet of the isolator, which is used to reduce the interaction between the inlet and the combustion chamber. The primary rocket is essentially a fuel-rich gas generator that acts as an igniter and flame holder for secondary fuel combustion. At the same time, the combustion chamber is designed to have an expanded shape to widen the operating range. This configuration contains complex physicochemical phenomena such as spray, supersonic turbulent combustion, etc. The sprayFoam solver in OpenFOAM16 is used to conduct simulation with the k-Omega SST turbulence model and PaSR combustion model. The inlet boundary conditions are enumerated in Table 1.

Table 2 Configuration of SOM-BPNN and test cases.

| Case No. | | SOM neurons/subdomain | Q error | BPNN topology | Testing error |
|---|---|---|---|---|---|
| A0 | A_ODE | - | - | - | - |
| A1 | A_100_31 | 30 × 30/225 | 0.07406 | 43-43 | $7.1 \times 10^{-04}$ |
| A2 | A_100_31-31 | | | 43-43-43 | $2.6 \times 10^{-04}$ |
| A3 | A_25_31 | 40 × 40/400 | 0.06068 | 43-43 | $6.7 \times 10^{-05}$ |
| A4 | A_25_31-31 | | | 43-43-43 | $3.7 \times 10^{-05}$ |
| A5 | A_25_31 | 50 × 50/625 | 0.05265 | 43-43 | $2.4 \times 10^{-05}$ |
| A6 | A_25_31-31 | | | 43-43-43 | $2.3 \times 10^{-05}$ |

The chemical alterations, including the species mass fractions, temperature, and pressure (for a total of 43 dimensions in this case), were recorded throughout the simulation to generate the entire data set. As for the performance of SOM, three different topologies were tested with 30 × 30 neurons, 40 × 40 neurons and 50 × 50 neurons, respectively. Using the trained SOM, the dataset was divided into subsets, and then each subset was assigned a BPNN for representation. Two alternative topologies are carried out and listed in Table 2. Combined with the SOM scheme, a total of 6 cases were finally implemented. It is worth noting that choosing the parameters of SOM (the neuron number) and BPNN (the number of hidden layers and neurons per layer) is a trade-off. Because an overly complex topology can lead to overfitting and more computational costs, while oversimplified one can result in low performance. Finally, the Q error and the mean testing error, which are the parameters to measure the training accuracy of SOM and BPNN, respectively, are also given in Table 2. The mean testing error was calculated by Eq. (3)

$$RMSE = \sqrt{\frac{1}{N}\sum_{i=1}^{N}(y_i^{exp} - y_i^{pre})^2} \tag{3}$$

where $N$ represents the total number of training data points, $y_i^{pre}$ is the $i$-th predicted value of the SOM-BPNN network, $y_i^{exp}$ are the actual value.

## 4. Results and discussion

In this part, the models developed above were applied to the RBCC mentioned in the previous section and the results are then discussed. unlike the training samples, large eddy simulation with 12 million structured cells is performed. First, as shown in Fig. 4, the wall pressure distributions of all cases are validated using experimental data, which was collected in a direct connect supersonic combustion test platform assembled in Science and Technology on Combustion, Internal Flow and Thermal-structure Laboratory (Northwestern Polytechnical University, Xi'an, China) [20]. The trend of the pressure distributions is consistent with the experimental results, indicating that the model used can simulate real working conditions and that different ANN topologies have slightly influenced on pressure prediction.

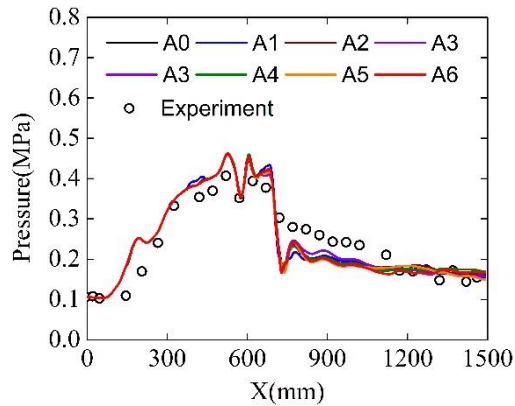

**Fig. 4.** Side surface pressures of simulation (cases A0-A5) and experiment [20].

Furthermore, Fig. 5 shows a comparison of instantaneous temperature and mass fraction distributions of selected species in cases A0 and A2. Simulations using the original chemistry and the ANN-based chemistry both capture the same flame structures as well as primary combustion (guided by primary rocket) and secondary combustion (triggered by the secondary fuel). Meanwhile, the ANN-based mechanism can represent the mass distribution of intermediate species and characterize local ignition and quenching caused by turbulent fluctuations. The simulation of these phenomena is highly demanding because, for supersonic combustion, a small error can cause large disturbances. For example, the ignition delay time calculated by the mechanism will determine whether or not a stable and self-sustaining combustion can be formed in the supersonic flow. Remarkably, the ANN-based mechanism leads to the same result as the original model, indicating an excellent performance.

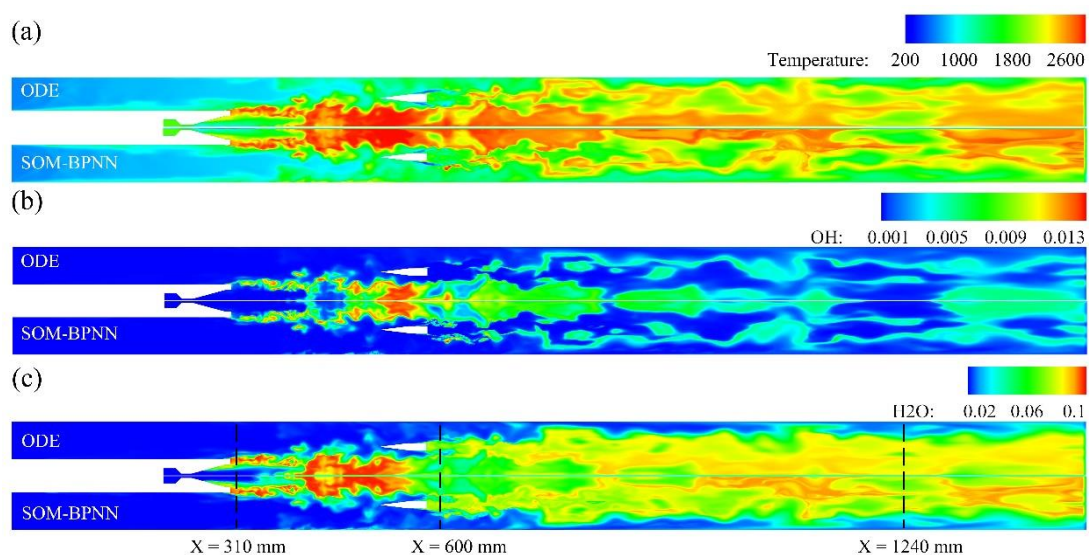

**Fig. 5.** Instantaneous temperature and mass fraction distributions of OH and $H_2O$ of Case A0 (top) and Case A1 (bottom).

To assess the accuracy of different topologies in more detail, time-averaged temperature and mass fraction profiles of selected species at three locations were further extracted (X = 310 mm, X = 600 mm and X = 1240 mm, marked in Fig. 5). These three locations represent different stages: 1) Combustion guided by the high temperature and fuel-rich gas flow exhausted from the primary rocket; 2) Mixing and intense combustion of secondary fuel injected by struts; 3) End of the combustion process. As shown in Fig. 6, in the first state, the curves show a sharp change, because the gas from the primary rocket is rich in reactive species (OH, $CH_2O$, etc.), which will start an intense reaction in an instant. Then, as the kerosene is injected and cracked, more reactants are added and burned, giving the double-peak curves (X = 600mm). One of the peaks is derived from the upstream reaction product, while the other comes from the injected kerosene. Eventually, as shown in the figures of X = 1240mm, oxygen cannot be mixed with the center stream in time due to the supersonic speed, so that some of the reactants are not consumed.

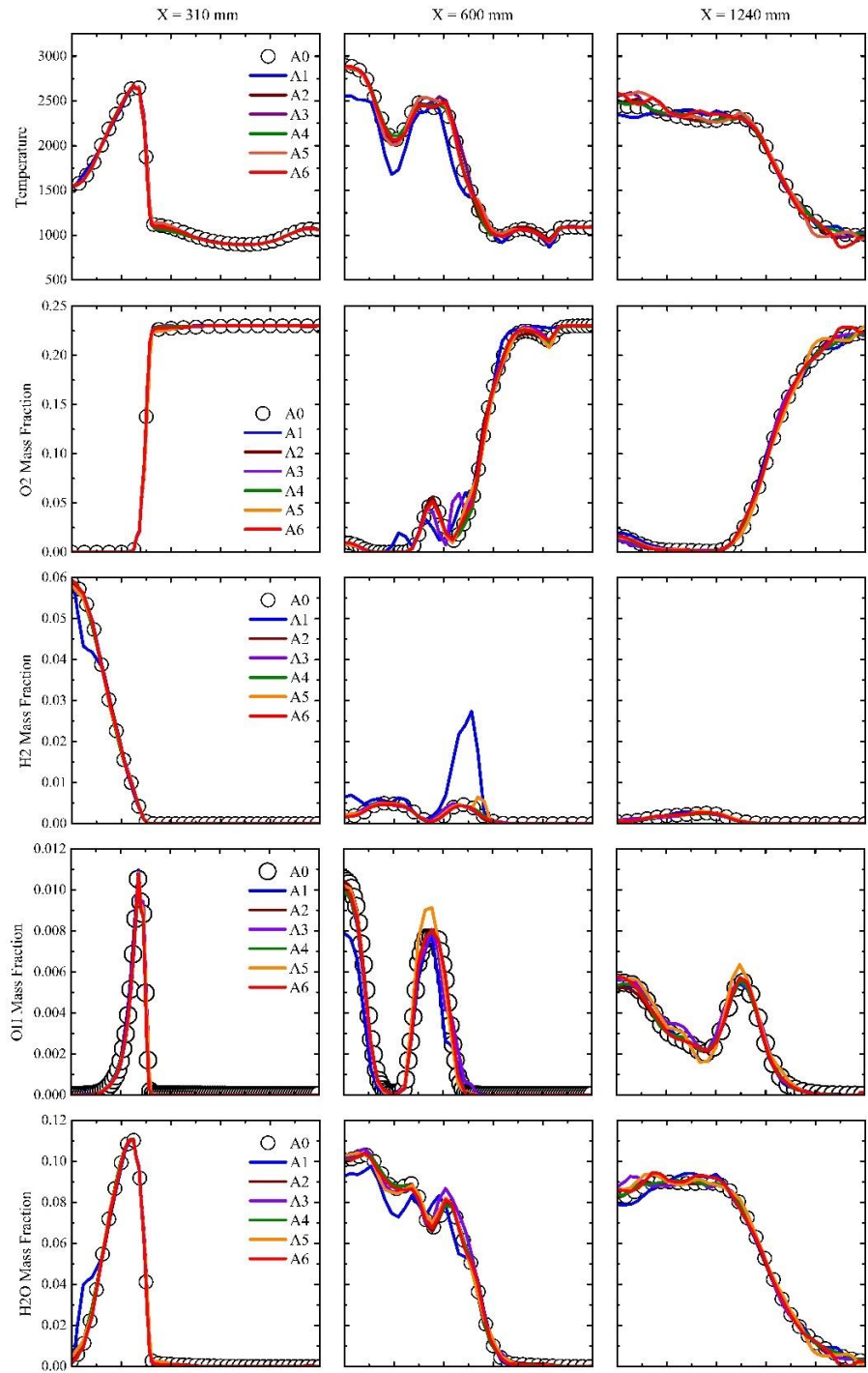

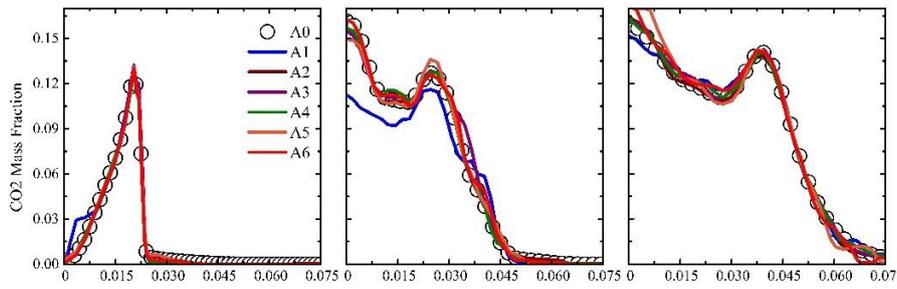

**Fig. 6.** Time-averaged temperature and mass fraction profiles of selected species at X = 310 mm, X = 600 mm and X = 1240 mm.

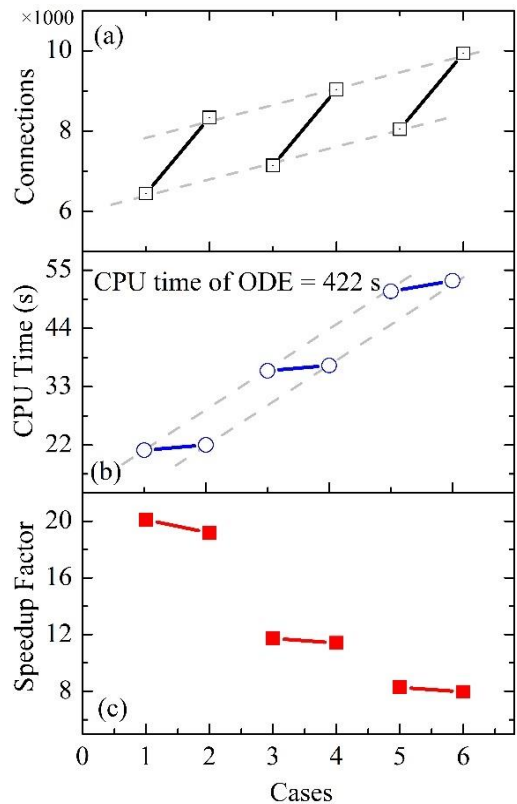

**Fig. 7.** (a) Number of connections, (b) CPU time and (c) Speedup factor of different cases.

The above results verified the accuracy of the new model, while its other performance metrics need to be demonstrated. Figure 7(a) shows the number of connections of different SOM-BPNN topologies. It represents the operations that the neural network needs to perform in calculating the chemical reactions. The results demonstrate that there is a linear relationship between the number of connections and the number of layers or the number of SOM neurons. Furthermore, combined with Fig. 7(b), the computational cost is also linearly related to the complexity of SOM-BPNN, while the calculation using ODE for reaction is proportional to the square of the species number. Finally, compared with solving 41 mechanisms directly, 8 to 20 times speedup is achieved by using different ANN topologies as shown in Fig. 7(c). These results encourage the application of this approach to more complex mechanisms.

## 5. Conclusions

In this study, we constructed an ANN-based chemical mechanism of kerosene with 41 species through a SOM-BPNN topology and verified it in a practical RBCC combustion chamber. Six topologies each with a different number of SOM neurons and BPNN layers were carried out to learn the dataset, which were generated by RANS simulations of the RBCC combustion chamber with less computational intensity. After training, the new modeling framework was applied to the LES simulation of the RBCC combustion chamber.

In terms of accuracy, five ANN-based models provide excellent consistency with the conventional ODE solver, while the case with the fewest neurons (Case A1) does not. As for efficiency, the CPU time of the ANN-based models was reduced typically by 8~20 times, compared with the conventional ODE solver with the 41 species skeleton mechanism of kerosene combustion. These results indicate that there is an optimal topology that can balance efficiency and accuracy, which is Case A2 in our study. Moreover, the computational cost of the ANN-based model is linearly proportional to the number of connections, while that of the ODE solver for reaction increases with the square of the species number, indicating a great potential of ANN-based models for more complex fuels.

**Acknowledgments**

This work was financially supported by the National Natural Science Foundation of China (Contract No. E060405). Support from the UK Engineering and Physical Sciences Research Council under the project "UK Consortium on Mesoscale Engineering Sciences (UKCOMES)" (Grant No. EP/R029598/1) and China Scholarship Council (CSC) is gratefully acknowledged.